\documentclass[12pt]{iopart}
\usepackage{graphicx}
\usepackage{multirow}
\usepackage{iopams}
\begin{document}
\title[]{Absolute absorption and dispersion of a rubidium vapour in the hyperfine Paschen-Back regime}
\author{Lee Weller, Kathrin S Kleinbach, Mark A Zentile, Charles S Adams and Ifan G Hughes}
\address{Joint Quantum Centre (JQC) Durham-Newcastle, Department of Physics, Rochester Building, Durham University, South Road, Durham, DH1 3LE, UK}
\author{Svenja Knappe}
\address{Time and Frequency Division, the National Institute of Standards and Technology, Boulder, Colorado 80305, USA}
\ead{lee.weller@durham.ac.uk}
\date{\today}

\begin{abstract}
\noindent 
Here we report on measurements  of the absolute absorption and dispersion properties of an isotopically pure $^{87}$Rb vapour for magnetic fields up to and including 0.6~T.  We discuss the various regimes that arise when the hyperfine and Zeeman interactions have different magnitudes, and show that we enter the hyperfine Paschen-Back regime for fields greater than 0.33~T on the Rb D$_{2}$ line.  The experiment uses a compact 1~mm$^{3}$ microfabricated vapour cell that makes it easy to maintain a uniform and large magnetic field with a small and inexpensive magnet.  We find excellent agreement between the experimental results and numerical calculations of the weak probe susceptibility where the line positions and strengths are calculated by matrix diagonalization.
\end{abstract}
\maketitle
\section{Introduction}
\label{Introduction}
The interaction of light with atomic ensembles continues to be a topic of active research.  Achieving strong coherent matter-light interfaces allows us to investigate applications within quantum information processing, such as the quantum internet~\cite{kimble2008quantum} and quantum memories~\cite{julsgaard2004experimental}.  The addition of an external magnetic field for such an ensemble further increases the possibilities for these interfaces to be realized.  Resonant and off-resonant linear and nonlinear magneto-optical effects in multi-particle ensembles have  allowed the observation of quantum teleportation between light and matter~\cite{sherson2006quantum} and control of atomic Zeeman populations in long-lived quantum memories~\cite{jenkins2012situ}.

As first observed by Zeeman in 1896~\cite{zeeman1897zeeman}, the energy levels and transition probabilities of an atomic ensemble are extremely sensitive to an external magnetic field.  Therefore a theoretical model for the absorption and dispersive properties of such ensembles has found utility in many applications, for example, Faraday dichroic beam splitter for Raman light~\cite{abel2009faraday}; Gigahertz-bandwidth atomic probes~\cite{siddons2009gigahertz}; Hanle-type coherent population trapping~\cite{noh2010calculated}; off-resonance laser frequency stabilization~\cite{marchant2011off}; realizing narrowband atomic filters~\cite{beduini2011ultra}; cooperative effects in an atomic nanolayer~\cite{keaveney2012cooperative}; imaging microwave fields in vapour cells~\cite{bohi2012simple} and achieving a compact optical isolator~\cite{weller2012optical}.     

The absorption and dispersion of an atomic ensemble in an external magnetic field can be calculated from the atomic susceptibility.  In the absence of field, absolute Doppler-broadened absorption~\cite{siddons2008absolute,kemp2011analytical} and dispersion~\cite{siddons2009off} in the low density regime, and dipole-dipole interactions~\cite{weller2011absolute} in the binary-collision regime have been tested.  In the presence of field, Stokes parameters for fields up to 0.08~T~\cite{weller2012measuring} have also been investigated.  The motivation for this work is to test the model for absolute susceptibility in a $^{87}$Rb vapour on the D$_{2}$ line.  To the best of our knowledge this is the first study highlighting excellent agreement between experimental results and numerical calculations for absolute absorption and dispersion for fields up to and including 0.6~T.  It should be noted that at such fields the nuclear spin, $I$, and total electronic angular momentum, $J$, are decoupled for both ground and excited states and the total angular momentum, $F$, is no longer a good quantum number; this is known as the hyperfine Paschen-Back regime~\cite{sargsyan2012hyperfine}. 

The structure of the rest of the paper is as follows: in Section~\ref{Theoretical Considerations} we discuss the quadratic magnetic term for fields encountered in the laboratory for low- and high-lying $n$ states and also the different regimes available for the linear magnetic term as a function of magnetic field.  In Section~\ref{Experimental Details} we describe the isotopically pure $^{87}$Rb microfabricated vapour cell and permanent neodymium magnet used in the investigation.  In Section~\ref{Absorption in the hyperfine Paschen-Back regime} we measure the evolution of the absolute optical depths as a function of field and detuning.  To highlight the decoupling we also show the transition-strength dependence for the outermost weakly allowed transitions.  In Section~\ref{Dispersion in the hyperfine Paschen-Back regime} we compare theory and experiment for the medium's absolute absorption and dispersive properties in the hyperfine Paschen-Back regime, highlighting the excellent agreement.  Finally, in Section~\ref{Conclusions}, we draw our conclusions.
\section{Theoretical Considerations}
\label{Theoretical Considerations}
\subsection{Atomic Hamiltonian}
\label{Atomic Hamiltonian}
The atomic Hamiltonian can be written as
\begin{eqnarray}
\hat{H} = \hat{H}_0 + \hat{H}_{\rm fs} + \hat{H}_{\rm hfs} + \hat{H}_{\rm Z}~,
\label{eq:hamilton}
\end{eqnarray}
where $\hat{H}_0$ is the coarse atomic structure; $\hat{H}_{\rm fs}$ and $\hat{H}_{\rm hfs}$ describes the fine and hyperfine interactions and $\hat{H}_{\rm Z}$; represents the atomic interaction with an external magnetic field.  The zero detuning frequencies in the absence of hyperfine splitting for $^{87}$Rb and $^{85}$Rb on the D$_{1}$ ($5^{2}S_{1/2} \rightarrow 5^{2}P_{1/2}$) and D$_{2}$ ($5^{2}S_{1/2} \rightarrow 5^{2}P_{3/2}$) lines are 377.11~THz~\cite{banerjee2004absolute} and 384.23~THz~\cite{arimondo1977experimental}, respectively.  The splittings associated with the hyperfine interaction Hamiltonian around the zero-detuning energies can be calculated by use of the following expression 
\begin{eqnarray}
\Delta E_{\rm hfs}= \frac{A_{\rm hfs}}{2} K 
                  + \frac{B_{\rm hfs}}{4} \frac{\frac{3}{2} K(K+1) - 2I(I+1)J(J+1)}{I(2I-1)J(2J-1)}~, 
\end{eqnarray}
where $A_{\rm hfs}$ is the magnetic dipole constant, $B_{\rm hfs}$ is the electric quadrupole constant and $K = F(F + 1) - I(I + 1) - J(J + 1)$; see equation 9.60 in~\cite{woodgate1983elementary}.  The numerical values to this expression can be found in, for example, table 1 of~\cite{siddons2008absolute}.  For an external magnetic field the magnetic interaction Hamiltonian has a linear and quadratic term in $\boldsymbol{\cal B}$~\cite{QM}.  For typical laboratory magnetic fields, quadratic shifts for states with a low principal quantum number, $n$, are extremely difficult to observe.  Therefore the quadratic term in $\boldsymbol{\cal B}$ can usually be ignored; however, with high $n$ states the dependence is very much evident~\footnote{The study of diamagnetic shifts in Rydberg atoms has been studied for many years.  For magnetic fields of about 0.6~T the shifts become evident for states above $n~\approx~36$~\cite{RA}.}.  The magnetic interaction Hamiltonian for an external field has the form
\begin{eqnarray}
\hat{H}_{\rm Z} = - (\boldsymbol{\mu}_I + \boldsymbol{\mu}_L + \boldsymbol{\mu}_S) \cdot \boldsymbol{\cal B}~, 
\end{eqnarray}
where $\boldsymbol{\mu}_I$, $\boldsymbol{\mu}_L$ and $\boldsymbol{\mu}_S$ are the magnetic moments of the nucleus, the orbital motion due to the electron, and the spin of the electron, respectively.  We can ignore the contribution due to the magnetic moment of the nucleus, $\boldsymbol{\mu}_I$, because the Bohr magneton is three orders of magnitude larger than the nuclear magneton~\cite{QM}.  The different spacings $\Delta E_{\rm Z}$ associated with the magnetic interaction Hamiltonian are discussed in section~\ref{Various magnetic regimes}.  In this work we investigate two cases: the hyperfine linear Zeeman regime (HLZ), where the magnetic interaction is treated as a perturbation to the hyperfine interaction, $\Delta E_{\rm Z} < \Delta E_{\rm hfs}$; and the hyperfine Paschen-Back regime (HPB) where the magnetic interaction is larger than the hyperfine interaction, yet smaller than the fine interaction, $\Delta E_{\rm hfs} < \Delta E_{\rm Z}$.  For even larger fields the magnetic interaction dominates both the fine and hyperfine interaction; this is known as the fine Paschen-Back regime (FPB).   
\subsection{Various magnetic regimes}
\label{Various magnetic regimes}
\begin{figure*}[t]
\centering
\includegraphics*[width=0.9\textwidth]{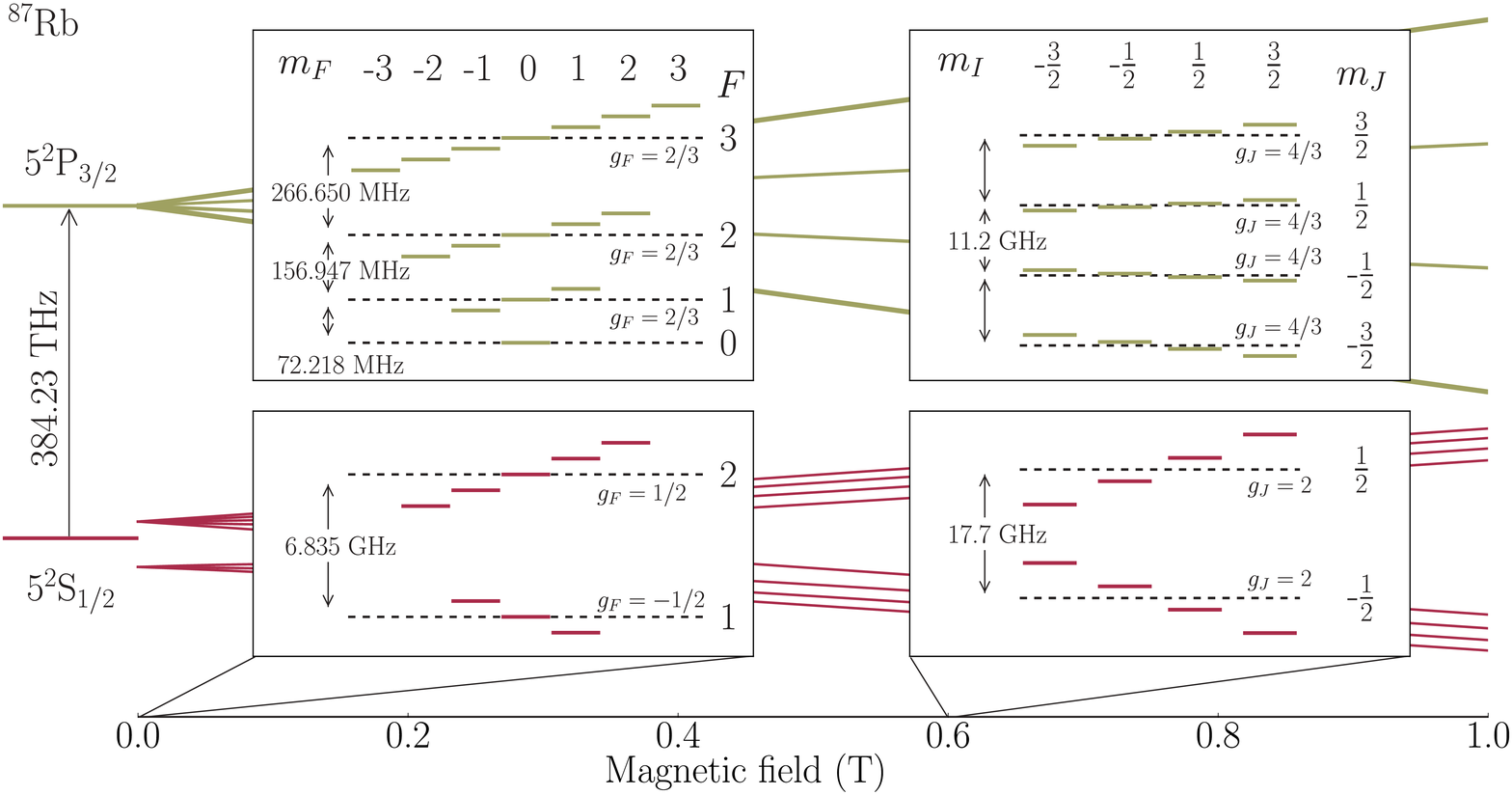}
\caption{Diagram showing the shift in the energy levels as a function of magnetic field for the $5^{2}S_{1/2}$ and $5^{2}P_{3/2}$ terms in $^{87}$Rb.  In the (weak-field) hyperfine linear Zeeman (HLZ) regime, the energy levels can be described by the $F$ and $m_{F}$ quantum numbers (shown in the left plot).  At intermediate fields there are no good quantum numbers to label all of the energy levels.  At large fields (0.6~T) $m_{I}$ and $m_{J}$ become good quantum numbers due to the nuclei and electronic spins decoupling (shown in the right plot); this case is referred to as the hyperfine Paschen-Back (HPB) regime.}
\label{Figure1}
\end{figure*}

Figure~\ref{Figure1} shows the energy level evolution as a function of magnetic field for $^{87}$Rb ($I = 3/2$) on the D$_{2}$ line.  In the weak-field regime the nuclear spin, $I$, and total electronic angular momentum, $J$, couple to give the total angular momentum, $F$, which has $2J + 1$ values for $J \leq I$.  For the $5^{2}S_{1/2}$ term of $^{87}$Rb, $F$ can be 1 or 2 with a hyperfine splitting of 6.8~GHz, whereas for the $5^{2}P_{3/2}$ term, $F$ can be 0,~1,~2 or 3 with hyperfine splittings between 70 and 270~MHz.  For weak magnetic fields each hyperfine level, $F$, is split into $2F + 1$ levels $(m_{F})$ symmetrically about the zero field level, $\Delta E_{\rm Z} = g_{F} m_{F} \mu_{B} B$; this is known as the HLZ regime.  In this regime the $\left|F,m_{F}\right\rangle$ basis best describes the interaction.  For large magnetic fields the total angular momentum decouples into $I$ and $J$; this is known as the HPB regime.  The effect introduces $2I + 1$ levels $(m_{I})$ with each $m_{J}$ value, for the $5^{2}S_{1/2}$ term, $m_{J}$ is equal to $1/2$ or $-1/2$ and for the $5^{2}P_{3/2}$ term, $m_{J}$ can be $3/2$,~$1/2$,~$-1/2$ or $-3/2$.  In this regime the $\left|m_{J},m_{I}\right\rangle$ basis best describes the interaction.  The spacings of the levels are proportional to the values of $m_{J}$ and $m_{I}$, $\Delta E_{\rm Z} = (g_{J} m_{J} + g_{I} m_{I}) \mu_{B} B$.  A detailed theoretical discussion of this regime can be found in~\cite{QM}.  For intermediate magnetic fields all symmetry is lost and there is no good basis to describe the interaction.  For even larger fields, typically $>218~$T, $J$ decouples into the total orbital angular momentum, $L$, and total spin angular momentum, $S$; this is known as the FPB regime.  This effect introduces $2L + 1$ levels $(m_{L})$ for each of the 2 orientations of $m_{S}$ for a single electron.  These are known as Lorentz triplets when $L = 1$.  The spacings of the levels are proportional to the values of $m_{L}$ and $m_{S}$, $\Delta E_{\rm Z} = (g_{L} m_{L} + g_{S} m_{S}) \mu_{B} B$. 

\begin{table*}[t]
\centering
\caption{The magnetic fields required to gain access to the linear and quadratic terms of the magnetic interaction Hamiltonian for Rb.  For a field of 0.6~T and $n = 5$ state only the hyperfine linear Zeeman (HLZ) and hyperfine Paschen-Back (HPB) regimes are accessible in this work.  The magnetic dipole constants, $A_{\rm hfs}$, and electric quadrupole constants, $B_{\rm hfs}$, were obtained from~\cite{arimondo1977experimental}.  The fine Paschen-Back (FPB) regime is also shown.  The $n \approx 36$ state would need to be investigated to see any quadratic effects for the field under investigation, or a field of $\approx~1.6\times 10^3$~T would be required to investigate the $n = 5$ state.}
\begin{tabular}{|c|c|ccc|c|c|}
\hline
\multirow{3}{*}{Isotope}    & \multirow{3}{*}{Term} &                     \multicolumn{4}{|c|}{Linear}                 & \multirow{2}{*}{Quadratic}    \\\cline{3-6}
           								  & 		  &  \multicolumn{3}{|c|}{HLZ / HPB}                          & \multicolumn{1}{|c|}{FPB}    &                       \\\cline{3-7}	
           								  & 		  & $A_{\rm hfs}$/h (MHz) & $B_{\rm hfs}$/h (MHz)    & B (T)  & B (T)                & B (T)                          \\\hline	
\multirow{3}{*}{$^{87}$Rb}  & 5$^{2}$S$_{1/2}$ & 3417.34           & -                 & 0.33   & -				             & \multirow{6}{*}{$\approx1.6\times 10^3$} \\ 
                    			  & 5$^{2}$P$_{1/2}$ & 407.24	           & -                 & 0.12   & \multirow{2}{*}{218} &	   		                        \\
     										    & 5$^{2}$P$_{3/2}$ & 84.72	           & 12.50             & 0.01	  &    		               &                                \\\cline{1-6}
\multirow{3}{*}{$^{85}$Rb}  & 5$^{2}$S$_{1/2}$ & 1011.91	         & -                 & 0.13   & -				             &   				                      \\
													  & 5$^{2}$P$_{1/2}$ & 120.32	           & -                 & 0.05   & \multirow{2}{*}{218} &  										          \\
													  & 5$^{2}$P$_{3/2}$ & 25.00	           & 25.79             & 0.005  &					             &   			 			  	              \\\hline
														& $n$ $\approx$ 36 & 	                 &                   &        &					             & 0.6   			 			  	          \\\hline
\end{tabular}
\label{Table1}
\end{table*}  

Table~\ref{Table1} shows the typical calculated magnetic fields to gain access to the HLZ, HPB and FPB regimes of the linear magnetic interaction term of $^{87}$Rb and $^{85}$Rb.  In addition, the fields and principal quantum number required to see any quadratic effects are noted.  To calculate the fields we equate the hyperfine energy splitting to the sum of the Zeeman shift of the lowest $m_F$ state of the upper $F$ value and the highest $m_F$ state of the lower $F$ value. The HLZ and HPB regimes are accessed for fields much smaller and bigger than the calculated values, respectively.  A similar procedure is adapted to calculate the fields required to access the FPB regime.  For the low-lying $n = 5$ state, fields of $\approx~1.6\times 10^3$~T are required to observe the quadratic term of the magnetic interaction Hamiltonian.  For 0.6~T we would need to gain access to the $n~\approx~36$ state to measure any such interaction.  
\subsection{Matrix representation of the Hamiltonian}
\label{Matrix representation of the Hamiltonian}
A detailed description of the model used to calculate atomic susceptibility incorporating the energy levels and transition probabilities of the Rb ensemble can be found in~\cite{weller2012measuring}.  In summary the lineshape around resonance is given by a convolution of the Lorentzian (accounting for natural, self-broadening and buffer gases) and a Gaussian distribution incorporating the Doppler shift due to thermal motion.  The total susceptibility, $\chi$, is then calculated by summing over the electric-dipole-allowed transitions.  From the atomic susceptibility we are able to model the absorptive and dispersive properties of the medium.  The refractive index and absorption coefficients can be calculated by use of the real, $n = 1 + \Re(\chi)/2$, and imaginary, $\alpha = k \Im(\chi)$, parts of the susceptibility, respectively, where $k$ is the wavevector.  To obtain numerical values for the susceptibility a matrix representation of the Hamiltonian for the fine and hyperfine interactions and atomic interaction with an external magnetic field are calculated in the completely uncoupled $\left|m_{I},m_{l},m_{s}\right\rangle$ basis.  The frequency detunings and transition strengths are calculated from a numerical diagonalisation of the matrix.    
\section{Experimental Details}
\label{Experimental Details}
The experimental procedure used to investigate the HPB regime is described in~\cite{weller2012optical}.  In summary, we use a linear polarized weak-beam~\cite{sherlock2009weak} with a power of 10~nW and $1/e^2$ radius of $80~\mu$m propagating along the $z$-direction.  The laser traverses a microfabricated cell of length $L=1$~mm.  The frequency axis is calibrated by use of the method described in~\cite{weller2012measuring}; for detunings over 60~GHz, two scans were stitched together in this experiment.  Transmission spectra are measured using a single calibrated photodiode.  Dispersion spectra use a balanced polarimeter to measure the light intensities of the horizontal, $I_{x}$, and vertical, $I_{y}$, channels after a polarization beam splitter.  A half-wave plate is set such that in the absence of rotation both channels of light are equal~\cite{huard1997polarization}.  For this publication we describe in detail the two main experimental components: the permanent neodymium magnet and $^{87}$Rb microfabricated vapour cell. 
\subsection{Permanent neodymium magnet}
\label{Permanent neodymium magnet}
\begin{figure}[t]
\centering
\includegraphics*[width=0.45\textwidth]{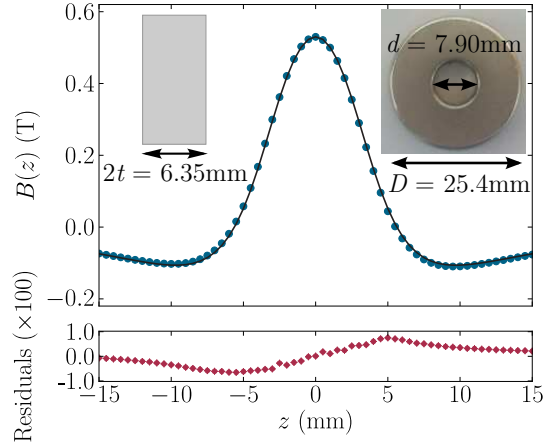}
\caption{Dimensions of a permanent neodymium magnet and axial variation of the $z$-component of magnetic field.  The length, $2t$, inner diameter, $d$, and outer diameter, $D$, describe the dimensions of the magnet, and the remanence, $B_{r}$, characterizes the strength of the material.  The measured solid (blue) circles are achieved by use of a Hall probe, and the theoretical solid (black) line is obtained from equation~$\ref{Bfield}$.  Below the main graph is a plot of the residuals (solid red circles), that show excellent agreement between theory and experiment, with an rms deviation of $0.4$~mT.  From a Marquardt-Levenberg fit, the parameters were found to be: $2t$ = (6.18 $\pm$ 0.12)~mm, $d$ = (7.98 $\pm$ 0.10)~mm, $D$ = (25.0 $\pm$ 0.6)~mm and $B_{r}$ = (1.42 $\pm$ 0.07)~T.}
\label{Figure2}
\end{figure}

To gain access to magnetic fields to investigate the HPB regime, an axial magnetized annular permanent neodymium magnet with a circular bore was chosen.  The direction of the beam in this experiment is parallel to the $z$-component of the magnetic field.  There is an analytic solution for the axial field of a uniformly magnetized annular magnet, which is~\cite{wynands1992compact} 
\begin{eqnarray}
B(z)&=\frac{B_r}{2}\left(\frac{z+t}{\sqrt{(z+t)^2+R^2}}-\frac{z-t}{\sqrt{(z-t)^2+R^2}}\right) \nonumber \\
		&-\frac{B_r}{2}\left(\frac{z+t}{\sqrt{(z+t)^2+r^2}}-\frac{z-t}{\sqrt{(z-t)^2+r^2}}\right)~, 
\label{Bfield}  
\end{eqnarray}
where $2t$ is the length, $d = 2r$ is the inner diameter of the magnet, $D = 2R$ is the outer diameter and $B_{r}$ is the remanence of the magnetic material.  Figure~\ref{Figure2} shows the measured solid (blue) circles and theoretical solid (black) line comparison for the axial variation of the $z$-component of magnetic field.  The error on the measured field and position is less than the size of the data point. Values of $2t$ = (6.18 $\pm$ 0.12)~mm, $d$ = (7.98 $\pm$ 0.10)~mm and $D$ = (25.0 $\pm$ 0.6)~mm were extracted from a Marquardt-Levenberg fit~\cite{MATU} by allowing the dimensions of the magnet to be free parameters.  The parameters are consistent with the physical dimensions of the magnet, with the deviation between theory and experiment at the $0.4$~mT rms level.  The excellent agreement validates the assumption of a uniform magnetization for the magnet.  The best-fit remanence for this magnet is $B_{r}$ = (1.42 $\pm$ 0.07)~T.    

\subsection{$^{87}$Rb microfabricated vapour cell}
\begin{figure}[t]
\centering
\includegraphics*[width=0.45\textwidth]{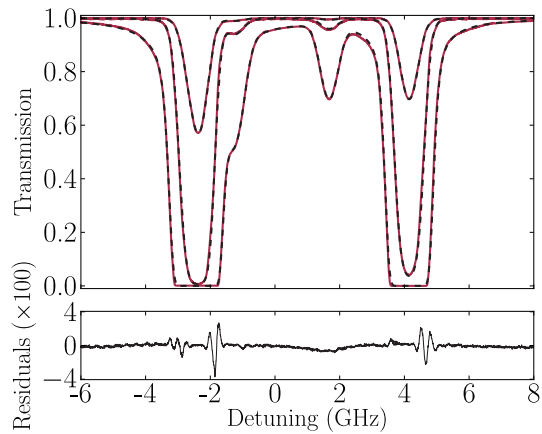}
\caption{Transmission plots for comparison between experiment and theory for the Rb D$_{2}$ line, through a 1~mm vapour cell (99$\%$ $^{87}$Rb, 1$\%$ $^{85}$Rb) as a function of linear detuning, $\Delta$/2$\pi$, for three different temperatures.  The solid (red) and dashed (black) lines show measured and expected transmission, respectively.  The temperatures of the vapours were extracted from a Marquardt-Levenberg fit~\cite{MATU} and were found to be (60.4 $\pm$ 0.2)~$^{\circ}$C (top), (90.1 $\pm$ 0.5)~$^{\circ}$C (middle) and (127.4 $\pm$ 0.8)~$^{\circ}$C (bottom).  The broadening and line shift due to the buffer gases were found to be $\Gamma_\mathrm{buffer}$/2$\pi$ = (23.7 $\pm$ 1.2)~MHz and $\Delta_\mathrm{buffer}$/2$\pi$ = (-7.9 $\pm$ 2.1)~MHz, for all three temperatures, respectively.  Below the main figure is a plot of the residuals between experiment and theory for the middle measurement.  There is excellent agreement between the data and model, with an rms deviation of $0.5\%$.  There is, however, a small number of glitches due to the linearisation of the laser scan being inadequate.}
\label{Figure3}
\end{figure}

\label{$^{87}$Rb microfabricated vapour cell}
A $1~\times~1~\times~1$~mm$^3$ heated isotopically pure $^{87}$Rb microfabricated vapour cell~\cite{knappe2005atomic} was used that contained buffer gases including hydrogen (H$_{2}$) and methane (CH$_{4}$).  To obtain the optical depths required to compare theoretical and measured absorption spectra, the cell was placed in an oven allowing the number density to vary over several orders of magnitude.  Figure~\ref{Figure3} shows a plot of the transmission through the heated vapour cell on the Rb D$_{2}$ line.  The solid (red) and dashed (black) lines show the measured and theoretical transmission, respectively, for several temperatures and number densities.  Based on the discussion in~\cite{weller2011absolute} we have previously accounted for a number-density-dependent increase in the Lorentzian width due to dipole-dipole interactions; however, we must now also include the broadening and shift due to the buffer gases.  The collisional broadening and shift of the Rb D$_{1}$ and D$_{2}$ lines by rare gases have previously been measured~\cite{rotondaro1997collisional}.  For the Rb D$_{2}$ line and buffer gases H$_{2}$ and CH$_{4}$, the broadening, $\Gamma_\mathrm{buffer}$/2$\pi$, equals 26.4~MHz~Torr$^{-1}$, 26.2~MHz~Torr$^{-1}$ and the line shift, $\Delta_\mathrm{buffer}$/2$\pi$, equals $-$3.8~MHz~Torr$^{-1}$, $-$7.0~MHz~Torr$^{-1}$, respectively.  From the analysis of figure~\ref{Figure3} we learn that the cell contains $99\%$ $^{87}$Rb, $1\%$ $^{85}$Rb, and a total pressure of $\approx$~Torr of H$_{2}$ and CH$_{4}$.
 
Having characterized the magnet and the microfabricated cell we could then perform spectroscopy in the HPB regime.

\section{Absorption in the hyperfine Paschen-Back regime}
\label{Absorption in the hyperfine Paschen-Back regime}
\begin{figure*}[t]
\centering
\includegraphics*[width=0.9\textwidth]{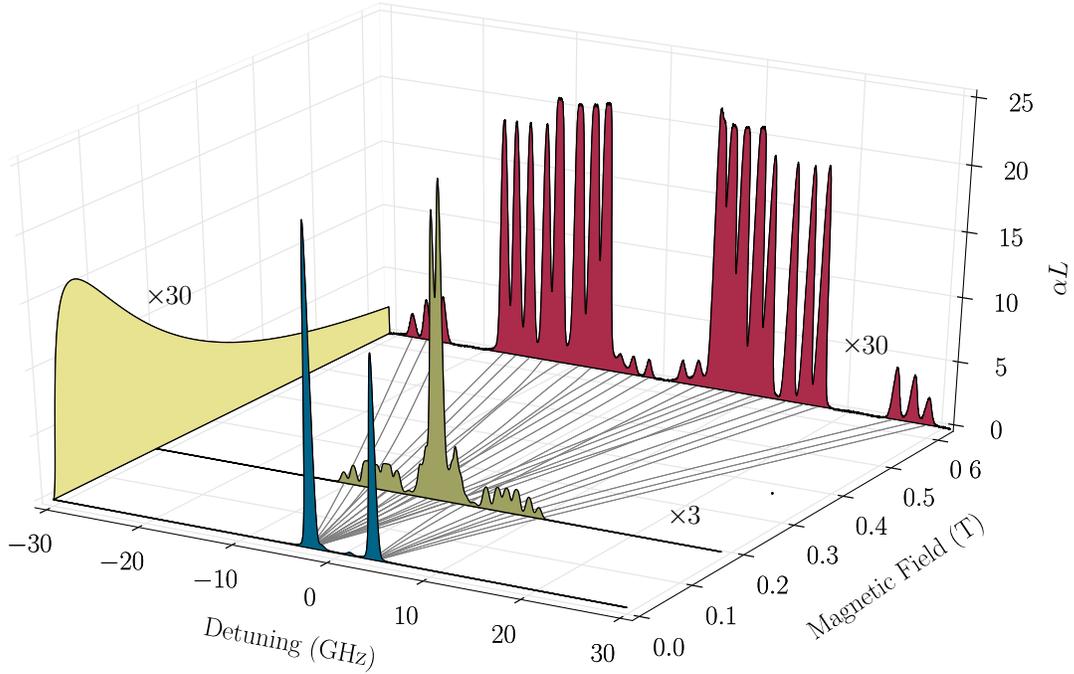}
\caption{Experimentally measured absolute optical depths for the Rb D$_{2}$ line, through a vapour cell (99$\%$ $^{87}$Rb, 1$\%$ $^{85}$Rb) of length, $L=1$~mm, as a function of linear detuning, $\Delta$/2$\pi$, at three different magnetic field values.  All three spectra were measured at a temperature of (116~$\pm$~1)~$^\circ$C.  The solid (blue) measured spectrum was taken in the absence of magnetic field.  The solid (olive) measured spectrum was taken at a field of (0.180~$\pm$~0.001)~T in the intermediate regime.  The solid (red) measured spectrum was measured at a field of (0.618~$\pm$~0.002)~T in the hyperfine Paschen-Back (HPB) regime.  The solid (grey) theoretical lines show the transition frequencies as a function of magnetic field.  Also plotted is the solid (yellow) theoretical transition strength of the outermost weak transitions as a function of magnetic field.  The normalisation factors ($\times3$ and $\times30$) compensate for a decrease in the transition strengths.}
\label{Figure4}
\end{figure*}

Figure~\ref{Figure4} shows the measured absolute optical depths as a function of detuning and magnetic field on the D$_{2}$ line in a $^{87}$Rb vapour.  The model for the susceptibility yields solid (grey) theoretical lines corresponding to the evolution of the transition frequencies as a function of magnetic field.  In the zero-field regime, the solid (blue) measured absolute optical depths are shown at a temperature of (116~$\pm$~1)~$^\circ$C, the two large features are the Doppler-broadened  $^{87}$Rb transitions from the $F=1$ and $F=2$ states. The other two very small features arise from the 1$\%$ of $^{85}$Rb in the cell.  In the intermediate regime, the solid (olive) measured absolute optical depths are shown at a temperature of (116~$\pm$~1)~$^\circ$C and for a field of (0.180~$\pm$~0.001)~T.  There are many spectral features; assigning quantum numbers to the transitions is difficult because there is no suitable basis set: the excited terms are completely decoupled, whereas the ground terms are only partially uncoupled. 

In the HPB regime, the solid (red) measured absolute optical depths are shown at a temperature of (116~$\pm$~1)~$^\circ$C and for a field of (0.618~$\pm$~0.002)~T.  The spectrum remains very rich in structure; however, the $\left|m_{J},m_{I}\right\rangle$ basis best describes the interaction, with the expected 16 strong transitions at this field being clearly visible.  Note that as the Zeeman shift exceeds the hyperfine interaction, the spectrum becomes symmetric with respect to detuning.  The other weaker transitions arise as a consequence of the ground state still not being completely decoupled.  For example, the second-highest energy state in the ground-level manifold asymptotes to being $\left|1/2,1/2\right\rangle$ at large field in the $\left|m_{J},m_{I}\right\rangle$ basis.  However, at 0.618~T the composition of the state is $0.99 \left|1/2,1/2\right\rangle + 0.14 \left|-1/2,3/2\right\rangle$.  The weak component of this state couples via an allowed $\Delta m_J = -1$, $\Delta m_I = 0$ transition to the lowest energy level of the excited manifold $\left|-3/2,3/2\right\rangle$.  The theoretical transition strength for the outermost of the weak transitions is shown as the solid (yellow) curve for all fields.  Numerically diagonalizing the atomic Hamiltonian matrix allows one to predict the energy levels and probabilities for all the features.  The 16 strong transitions ($\Delta m_J = \pm~1$) are easily obtained from figure~\ref{Figure1}; however, the advantage of our technique is that the detunings and absolute linestrengths of the 12 weakly allowed transitions are also given.  

\section{Dispersion in the hyperfine Paschen-Back regime}
\label{Dispersion in the hyperfine Paschen-Back regime}
\begin{figure*}[t]
\centering
\includegraphics*[width=0.9\textwidth]{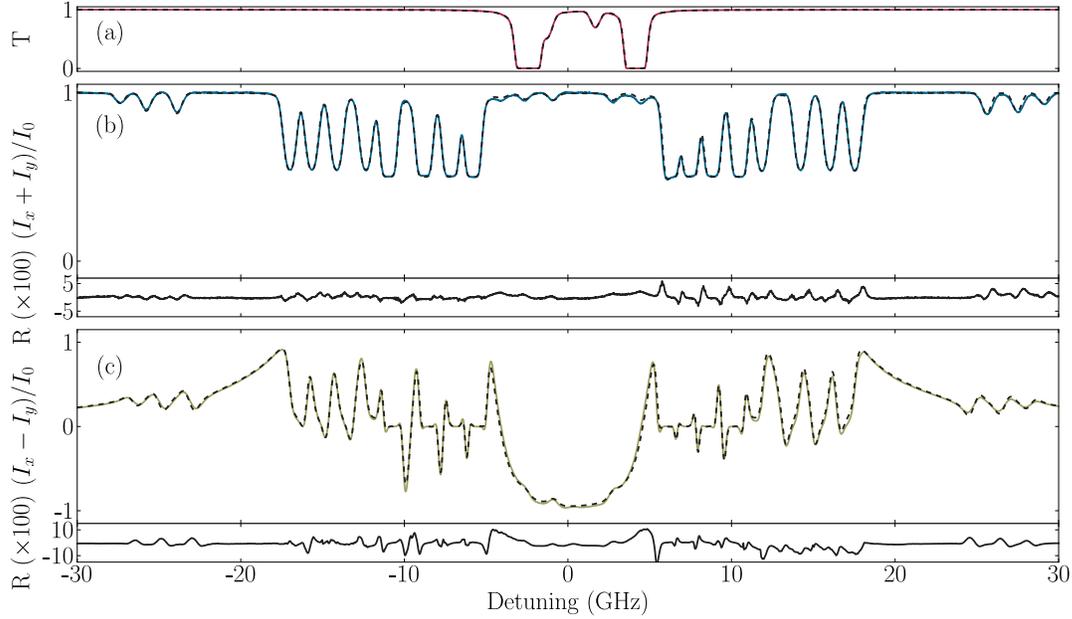}
\caption{Measured and theoretical absolute dispersion and absorption spectra in the hyperfine Paschen-Back (HPB) regime for the Rb D$_{2}$ line, through a vapour cell (99$\%$ $^{87}$Rb, 1$\%$ $^{85}$Rb) of length, $L=1$~mm, as a function of linear detuning, $\Delta$/2$\pi$.  Plot~(a) shows solid (red) measured and dashed (black) theoretical transmission in the absence of field at a temperature of (127.4 $\pm$ 0.8)~$^{\circ}$C and a width of $\Gamma_\mathrm{buffer}$/2$\pi$ = (23.7 $\pm$ 1.2)~MHz.  Plot~(b) shows solid (blue) measured and dashed (black) theoretical transmission in the presence of a field of (0.618~$\pm$~0.003)~T, at a temperature of (114.8 $\pm$ 0.6)~$^{\circ}$C and a width of $\Gamma_\mathrm{buffer}$/2$\pi$ = (47 $\pm$ 10)~MHz.  Below is a plot of residuals showing the excellent agreement between theory and experiment in plot~(b), with an rms of 0.9~$\%$.  Plot~(c) shows solid (olive) measured and dashed (black) theoretical differencing signals in the presence of a field of (0.599~$\pm$~0.003)~T at a temperature of (126 $\pm$ 0.8)~$^{\circ}$C and a width of $\Gamma_\mathrm{buffer}$/2$\pi$ = (63 $\pm$ 12)~MHz.  The bottom plot is the residuals for (c), with an rms of 3.3~$\%$.  There was no attempt made to add in the shift due to the buffer gases.}
\label{Figure5}
\end{figure*}

Figure~\ref{Figure5} shows the measured and theoretical absolute dispersion and absorption in the HPB regime as a function of detuning on the D$_{2}$ line in a $^{87}$Rb vapour.  Plot~(a) shows solid (red) measured and dashed (black) theoretical absolute transmission at a temperature of (127.4~$\pm$~0.8)~$^{\circ}$C, showing the Rb absorption features in the absence of a magnetic field.  Plot~(b) shows solid (blue) measured and dashed (black) theoretical absolute transmission at a temperature of (114.8~$\pm$~0.6)~$^{\circ}$C in the presence of a magnetic field of (0.618~$\pm$~0.003)~T.  The transmission signal describing the absorptive properties of the medium, was measured using a single photodiode after the cell.  Below the main plot are the residuals showing excellent agreement over 60~GHz, with an rms of 0.9~$\%$.  Plot~(c) shows the solid (olive) measured and dashed (black) theoretical absolute dispersion at a temperature of (126.0~$\pm$~0.8)~$^{\circ}$C in the presence of a magnetic field of (0.599~$\pm$~0.003)~T.  The differencing signal describing the dispersive properties of the medium, was measured using a balanced polarimeter.  Below the main plot, residuals show good agreement over 60~GHz, with an rms of 3.3~$\%$.  In the theoretical and measured signals the additional weak features are clearly visible, owing to the fact that the ground terms are not completely decoupled.  Such a result highlights the strength of a model for understanding the energy levels and transition probabilities of such an ensemble.  In contrast to the zero crossings associated with the dispersive features of the off-resonant Faraday effect~\cite{marchant2011off}, the various features in figure~\ref{Figure5} are associated with Zeeman shift resonances and are therefore less sensitive to temperature.  Consequently this opens up the possibility for locking far off-resonance. 
\section{Conclusions}
\label{Conclusions}
In summary, we have tested our model for the electric susceptibility of Rb vapour for magnetic fields up to and including 0.6~T, which corresponds to the HPB regime on the D$_{2}$ line.  We have demonstrated excellent agreement between the theoretical predictions and the experimental measurements of the absolute absorption and dispersion properties of the medium. Our study extends the range of magnetic fields for which the theoretical model has been tested by an order of magnitude compared to previous work.  Understanding the optical properties of atomic vapours in the HPB regime will find utility in many applications, such as realising compact optical isolators~\cite{weller2012optical}, measuring magnetic fields with submicron spatial resolution, and constructing tunable atomic frequency references~\cite{sargsyan2012hyperfine}.
\ack
This work is supported by the Engineering and Physical Sciences Research Council (EPSRC).    
  
\bibliographystyle{iopart-num}
\bibliography{myreferences}
\end{document}